\documentclass[class=preprint]{nature}
\usepackage{graphicx}
\graphicspath{{../../Figures/}}
\usepackage[usenames,dvipsnames]{color}
\usepackage{siunitx}
\usepackage[colorlinks=true,citecolor=blue,linkcolor=blue,urlcolor=blue]%
{hyperref}
\usepackage[a4paper]{geometry}
\usepackage{fixltx2e}
\usepackage{upgreek}
\usepackage{amsmath}
\usepackage{amssymb}
\usepackage{amsthm}
\usepackage{mathrsfs}
\usepackage{amsbsy}
\usepackage{amstext}
\usepackage{xcolor}
\usepackage{float}
\usepackage{multirow}
\usepackage{wasysym}
\usepackage{mathtools}
\usepackage{subfigure}
\usepackage{mhchem}
\usepackage{amsfonts}
\usepackage{dcolumn}
\usepackage{placeins}
\usepackage{float}
\usepackage{caption}
\usepackage[right]{lineno}
\usepackage{fancyhdr}
\usepackage{bbold,bm}
\usepackage{pdfpages}
\usepackage[utf8]{inputenc}
\usepackage[T1]{fontenc}%
\usepackage{setspace}
\providecommand{\U}[1]{\protect\rule{.1in}{.1in}}
\DeclareCaptionLabelSeparator{vline}{$\boldsymbol{:}$}
\newcommand{\red}[1]{\color{red}#1 \color{black}}

\renewcommand{\red}[1]{#1}

\title{Discovery of enhanced lattice dynamics in a  single-layered hybrid perovskite}
\author{Zhuquan Zhang$^{1\dagger}$, Jiahao Zhang$^{2\dagger}$, Zi-Jie Liu$^{1\dagger}$, Nabeel S. Dahod$^3$, Watcharaphol Paritmongkol$^{1,3}$, Niamh Brown$^{1,3}$, Yu-Che Chien$^1$, Zhenbang Dai$^2$, Keith A. Nelson$^{1*}$, William A. Tisdale$^{3}$, Andrew M. Rappe$^{2}$, Edoardo Baldini$^{4*}$ \\
\normalsize{$^1$Department of Chemistry, Massachusetts Institute of Technology, Cambridge, Massachusetts, USA, 02139 }\\
\normalsize{$^2$Department of Chemistry, University of Pennsylvania, Philadelphia, Pennsylvania, USA, 19104-6323}\\
\normalsize{$^3$Department of Chemical Engineering, Massachusetts Institute of Technology, Cambridge, Massachusetts, USA, 02139 }\\
\normalsize{$^4$Department of Physics, The University of Texas at Austin, Austin, Texas, USA, 78712 } \\
\normalsize{$^{*}$E-mail: edoardo.baldini@austin.utexas.edu, kanelson@mit.edu} \\
\normalsize{$^\dagger$These authors contributed equally to this work} \\
\\
}
\begin{document}

\maketitle

\newpage
\section*{Abstract}
Layered hybrid perovskites have attracted much attention in recent years due to their emergent physical properties and exceptional functional performances, but the coexistence of lattice order and structural disorder severely hinders our understanding of these materials. One unsolved problem regards how the lattice dynamics are affected by the dimensional engineering of the inorganic frameworks and the interaction with the molecular moieties. Here, we address this question by using a combination of high-resolution spontaneous Raman scattering, high field terahertz spectroscopy, and molecular dynamics simulations. This approach enables us to reveal the structural vibrations and disorder in and out of equilibrium and provides surprising observables that differentiate single- and double-layered perovskites. While no distinct vibrational coherence is observed in double-layer perovskites, we discover that an off-resonant terahertz pulse can selectively drive a long-lived coherent phonon mode through a two-photon process in the single-layered system. This difference highlights the dramatic change in the lattice environment as the dimension is reduced. The present findings pave the way for the ultrafast structural engineering of hybrid lattices as well as for developing high-speed optical modulators based on layered perovskites.

\newpage
\section*{Main Text}
Over the past decade, two-dimensional hybrid perovskites (2DHPs) have emerged as natural quantum-well-like semiconductors with marked light absorption, large luminescence quantum yield \cite{gong2018electron,grancini2019dimensional}, and strong exciton binding energy\cite{smith2019tuning,mauck2019excitons}. Unlike their 3D counterparts, 2DHPs also show wider chemical variability and structural diversity, as their composition can be tuned by altering organic spacer cations, inorganic networks, and the number of octahedral layers.\cite{li20212d,shi2018two,saparov2016organic,paritmongkol2019synthetic} This richness may also lead to a plethora of emergent properties including ferroelectricity\cite{hang2011metal,chen2019two}, spin selectivity\cite{kim2021chiral}, and multifunctionality\cite{li2017chemically}. Although substantial efforts have been made to exploit this versatility by dimensional tailoring, it is an ongoing task to establish the structure-function relationships in these materials. Key to this goal is an understanding of the interplay between the inorganic lattice framework and the organic cations.\cite{menahem2021strongly,mayers2018lattice} Previous mechanistic studies have suggested that the hybrid lattice features significant anharmonicity and polarizability, in conjunction with structural disorder\cite{egger2018remains,miyata2018ferroelectric,rivett2018long}. However, it remains an open question whether these properties persist to the single-layer limit when the octahedral framework does not fully resemble the lead-halide backbone of the bulk perovskite structure. 

Here, we present a joint experimental-theoretical study aimed at uncovering the origin of the dynamic structural complexity in 2DHPs. By means of steady-state and ultrafast spectroscopy experiments, we identify unique fingerprints that distinguish the structural dynamics of the hybrid lattices in the crossover between quasi-2D and 2D. We observe that the collective motion of the octahedral cages is significantly enhanced as the number of layers per repeating unit is altered from two to one. Our results are rationalized via molecular dynamics calculations, which provide an atomic-level understanding of the hybrid lattice in and out of equilibrium.

We focus on two prototypical 2DHPs, which differ in the number of corner-sharing octahedral layers n: (BA)$_2$PbBr$_4$ (single-layer, n=1) and (BA)$_2$MAPb$_2$Br$_7$ (double layer, n=2). As shown in Fig.1a, n=1 2DHP only contains organic spacer ligands (BA=butylammonium) that separate the octahedral layers, while the double-layered 2DHP consists of additional A-site cations (MA=methylammonium) that occupy cuboctahedral pockets formed by eight octahedra. 

As a first step in the study of the lattice dynamics in thermal equilibrium, we monitor low-energy collective responses and structural disorder via high-resolution spontaneous Raman scattering. We map the Raman spectra of both 2DHPs across a wide range of temperatures as shown in Fig. 1b (n=1) and Fig. 1c (n=2). For n=1 2DHP, the Raman spectra are characterized by eight well-defined peaks that include a dominant mode at 1.8 THz.\cite{dhanabalan2020directional} In contrast, the Raman data of n=2 2DHP exhibit broad features at all temperatures. In Fig. 1d, we provide a detailed comparison by selecting the Raman spectra of both materials at room temperature and at 77 K. The room temperature spectra for both n=1 and n=2 crystals show a prominent spectral continuum below 5 THz, which is reminiscent of quasi-elastic peaks due to disorder\cite{hlinka2008coexistence,yaffe2017local,dahod2020low}; however, the 1.8 THz peak in the n=1 compound stands out from the underlying continuum. When the temperature is decreased to 77 K, the continuum background contribution to the n=1 Raman response is dramatically reduced, and several additional modes become distinct. In contrast, the low-temperature spectrum of n=2 2DHP exhibits persistent broad features, suggesting the coexistence of phonon peaks and structural disorder. This is because that in the n=2 system, the presence of the MA cation introduces both dynamic and orientational disorder. From the steady-state Raman data, we establish a thermal equilibrium view of the structural complexity in 2DHPs and conclude that the structural disorder is significantly reduced in the single-layer limit. 

Next, we track the time evolution of lattice response at ultrafast timescales to attain more insights into the structural dynamics and separate the collective modes from the dynamic disorder\cite{yan1987impulsive,dougherty1992femtosecond}. To monitor the lattice behavior in real-time, we employ THz field-induced Kerr effect (TKE) spectroscopy\cite{hoffmann2009terahertz} (see Fig. 2a). Unlike previous studies that used optical pump pulses for Kerr effect spectroscopy\cite{zhong2008optical} to investigate the lattice dynamics\cite{miyata2017large}, molecular reorientation\cite{zhu2016screening}, and light propagation\cite{maehrlein2021decoding} in hybrid perovskites, we take advantage of the slowly varying electric field of THz pulses to induce giant polarizability responses\cite{melnikov2022terahertz} that may not be accessible with the pump pulses in the optical frequency range.

Figure 2b shows the TKE data recorded at several temperatures for both 2DHPs. For the n=2 sample at room temperature (green curve), only non-oscillatory signals are observed after the initial electronic response induced by the THz pulse. When the temperature is lowered below 160 K, oscillations emerge but only last for a few cycles, less than 5 ps. These observations indicate that in n=2 2DHP, any excited phonon mode loses its coherence quickly, consistent with the observation of broad features in the steady-state Raman spectra. In contrast, TKE signals from the n=1 2DHP exhibit a strikingly different behavior. First, the response is 10 times larger than its n=2 counterpart. This distinction indicates the presence of an enhanced polarizability at THz frequencies, as the two materials have similar dielectric properties at the probe photon energy (see Supplementary Note 1). Second, the initial response shows a bipolar character, signaled by the two lobes with opposite signs that do not follow the incident THz waveform. This type of signal has only been reported in aqueous systems where it has been attributed to different components of the liquid dynamics.\cite{elgabarty2020energy,zhao2020ultrafast} It is therefore surprising to observe such a response in a crystalline material, and we attribute its origin to the BA organic ligands and the unique response of the single-layer octahedral networks\cite{reyes2020unraveling} (see Supplementary Note 1). More importantly, we observe that a long-lived sinusoidal modulation appears promptly after the initial electronic response. The oscillation frequency of 1.8 THz (see Fig. 2c) corresponds to the most prominent phonon mode in the equilibrium Raman spectra. Even at room temperature, this coherent phonon response remains detectable for more than 10 ps. Decreasing the temperature results in a simultaneous increase in the oscillation amplitude and decay time (see Fig. 2d and 2e), which is a manifestation of the suppressed structural disorder and reduced anharmonic decay of the excited mode.

To identify the underlying mechanism that drives this long-lived coherent phonon response, we obtain the THz field-dependent TKE signals for n=1 2DHP at 10 K. As shown in Fig. 3a, the time-domain oscillatory responses increase monotonically with the THz field strengths. Fourier transformation of the oscillatory parts of the TKE signals shows that the spectral amplitude scales as the square of the pump electric field (see Fig. 3b). This observation indicates that the Raman mode at 1.8 THz is driven through a second-order interaction with the THz pump field. Such a nonlinear excitation process can proceed through two distinct pathways, which are ionic or photonic in nature (see Supplementary Note 2).\cite{khalsa2021ultrafast} In the ionic scenario, the THz field drives a dipole-active mode and nonlinear excitation of the Raman mode is mediated by anharmonic phonon-phonon coupling,\cite{forst_nonlinear_2011,juraschek2018sum} whereas in the photonic mechanism, the THz field drives the Raman mode directly through the nonlinear polarizability.\cite{aku2005long, johnson2019distinguishing, maehrlein2017terahertz} \red{Since the frequency of the Raman mode is well above the bandwidth of the incident THz pulse (see Fig. 2c), we can also exclude the scenario of impulsive stimulated Raman scattering\cite{aku2005long,yan1985impulsive} or impulsive ionic Raman scattering\cite{maradudin1970ionic, wallis1971ionic, humphreys1972ionic}, in which difference-frequency components of the photon or phonon field drives the Raman mode. Rather, the sum-frequency excitation pathway should be responsible for the observed Raman excitation. For such a process to be ionic, there must be an infrared-active phonon mode directly driven by the THz pump, whose phonon frequency is ideally at half of the Raman mode frequency (i.e., $\Omega_{IR}=\Omega_R/2\sim$ 0.9 THz). Since the crystal is centrosymmetric, only Raman-active modes produce transient birefringent signals, and therefore oscillations in the TKE responses do not reflect coherent excitation of infrared modes directly driven by the THz field.\cite{li2019terahertz,von2022amplification} To provide further clues, we apply time-domain THz spectroscopy, which directly measures dipole-allowed transitions in the THz frequency range. Figure 3c shows the imaginary part of dielectric permittivity at various temperatures below 200 K, which reveals two infrared-active modes (i.e., at 0.5 and 0.75 THz) emerging at low temperatures. There is no phonon mode at 0.9 THz that fulfills the resonance condition. Furthermore, it was previously established that the sum-frequency ionic Raman scattering would lead to beat signals in the time-domain, resulting from the mutual exchange of energy between the driven infrared-active and Raman-active phonon modes.\cite{juraschek2018sum} In our data, we only observe an exponential decay of the Raman coherence. This rules out any possibility that the 1.8 THz Raman-active mode is driven by anharmonic coupling to an excited infrared-active phonon mode. Therefore, we can conclude that the observed coherent collective response is generated by sum-frequency photonic excitation (see Fig. 3d).\cite{maehrlein2017terahertz,mead2020sum} However, we cannot exclude that the 0.75 THz infrared active phonon mode in the bandwidth of our THz field acts as a real intermediate state to our Raman-like process, especially at low temperature. This mechanism, which has been recently discovered theoretically and dubbed as infrared resonant Raman effect\cite{khalsa2021ultrafast}, depends on the ionic degrees of freedom but relies on the nonlinear lattice polarizability rather than on anharmonic phonon-phonon interactions. Such an effect may therefore explain why we observe a much more prominent Raman excitation response at low temperature.}

To gain deeper insight into the nature of both thermal and coherent dynamics, we conduct \emph{ab initio} molecular dynamics (MD) simulations for a $\sqrt{2} \times \sqrt{2} \times 2$ cell. To simulate equilibrium states, we use the canonical (NVT) ensemble for calculating the spontaneous Raman responses. Figure 4a displays the simulated Raman spectra for both n=1 and n=2 2DHPs at 77 K, which agree well with the experimental data at the same temperature. While the n=2 2DHP exhibits a more disorder-dominated Raman response, the Raman spectrum of the n=1 2DHP shows a distinct peak at around 1.9 THz, which is close to the frequency of the most prominent phonon mode observed in the experimental data (i.e., 1.8 THz). Based on real-space analysis of the MD simulations, we identify this mode as the bending and twisting of the octahedral cages in the single-layer inorganic framework. \red{Since in the 2DHP with n=2, the two adjacent layers of lead bromide octahedra are bonded to each other, the octahedral motions are significantly more coupled as compared to the n=1 system. The rotation and rattling motions of additional MA organic cations also introduce additional local static disorder, leading to much more spatially uncorrelated lattice dynamics associated with heterogeneous responses. This naturally explains why the Raman spectra show much broader features for the n=2 system.} We also investigate the nonlinear THz light-matter interaction in the n=1 2DHP. To capture coherent dynamics rather than the thermal fluctuation of the lattice response, we conduct MD simulations in the microcanonical ensemble (NVE). We apply the experimentally measured THz electric field waveform to our system, setting its polarization in the plane of the octahedral layers. We then project the MD trajectory into the eigenmode basis calculated from density functional perturbation theory (DFPT).\cite{gonze1995adiabatic,baroni1987green,baroni2001phonons} As shown in Fig. 4b, we find that the driving THz field generates a long-lived oscillatory response that does not decay to zero up to 20 ps. The Fourier-transform spectrum in Fig. 4c reveals a sharp peak centered at around 1.9 THz, in excellent agreement with our experimental data.

\section*{Discussion}
Our spectroscopic measurements combined with the MD simulations establish that the single-layered hybrid perovskite features a giant polarizable lattice response that does not exist in the double-layered perovskite counterparts. This finding highlights the use of tailored THz light excitation to study hybrid lattices exhibiting a complex interplay of molecular and ionic dynamics. From the fundamental point of view, this approach can be applied to explore many other structurally complex materials, including artificially engineered heterostructures and moiré superlattices, and opens the door to desirably controlling their emergent properties and novel functionalities with light.\cite{disa2021engineering} Given that our sample thickness is $\sim$ 100 $\mu$m, the estimated modulation depth of the THz field-induced polarization rotation is $\sim$ 2 dB/mm at room temperature (see Supplementary Note 4). Although we only demonstrate the polarization modulation for light of a single wavelength (800 nm), we expect that similar results will hold for a broad range of wavelengths below the material's bandgap (i.e., > 400 nm). For these reasons, we believe that 2DHPs are promising candidates for achieving all-optical, broadband refractive modulators at high speeds with tailored THz stimuli\cite{grinblat2019ultrafast}, offering new perspectives for the development of novel optical devices\cite{sun2016optical}.

\newpage

\section*{Methods}
\noindent \textit{Synthesis of bromide-lead 2DHPs} \\
Crystals of bromide-lead 2DHPs were synthesized by slow-cooling crystallization following a previously reported procedure\cite{paritmongkol2019synthetic}. Firstly, a solution of lead (II) bromide (PbBr$_2$) was prepared by dissolving PbO (99.9+$\%$, (trace metal basis) <10 microns, powder, ACROS Organic) in an aqueous hydrogen bromide solution (HBr, ACS reagent, 48$\%$, MilliporeSigma). Then, a small volume of butylamine (BA) was added to the PbBr$_2$ solution to form a white precipitate of (BA)$_2$PbBr$_4$ (single-layer, n=1). To prepare (BA)$_2$MAPb$_2$Br$_7$ (double layer, n=2), a solution of methylammonium bromide (MABr) salt was prepared in a separate vial by dissolving the salt in an aqueous HBr solution. The MABr solution was subsequently added into the solution of (BA)$_2$PbBr$_4$ to form a (BA)$_2$MAPb$_2$Br$_7$ solution. Next, the solutions of (BA)$_2$PbBr$_4$ and (BA)$_2$MAPb$_2$Br$_7$ were further diluted by additional volumes of HBr before being heated to 130 °C until they became clear. After that, they were allowed to cool slowly to room temperature inside a thermos filled with hot sand at 110 °C to induce crystallization. Crystals were then collected by suction filtration and dried under reduced pressure for at least 12 hours. The quantities of reagents used can be found in Table S1.

\noindent \textit{High-resolution spontaneous Raman scattering} \\
Steady-state Raman spectra were collected in a backscattered geometry using a home-built micro-Raman instrument. Samples were housed within an optical cryostat (Janis ST-500, fused quartz window) mounted to an inverted Nikon microscope (60x, 0.6 NA objective), and kept under vacuum during all measurements. A 785 nm narrow-band continuous wave excitation source was filtered from undesirable amplified spontaneous emission using a series of cleanup filters (laser and filters from Ondax). The Rayleigh line was minimized by passing the collected signal through a set of volume holographic grating notch filters (from Ondax) before being dispersed in a 0.5 m focal length spectrograph (SP-2500, Princeton Instruments) using a 1200 g/mm and 750 nm blaze grating. The resulting spectrum was imaged with a cooled charge-coupled device camera (Princeton Instruments Pixis) with a typical signal integration time of 15-30 s. The Rayleigh notch filters, centered at 0 cm$^{-1}$, have a full attenuation bandwidth of ± 10 cm$^{-1}$. The overall spectral resolution of the instrument is 0.9 cm$^{-1}$. The spectra were calibrated by comparison to the longitudinal optical phonon position of a CdSe standard.

\noindent \textit{TKE spectroscopy}\\
\noindent The majority of the output of a 1 kHz Ti:Sapphire laser amplifier (Coherent Legend Elite Duo, 800nm, 12 mJ, 35 fs) was chopped at 500 Hz and used to generate single-cycle THz pulses via optical rectification process with a tilted pulse front\cite{yeh2008generation}. The THz pulse was collected and focused by a pair of 90° off-axis parabolic mirrors. The remainder of the laser output was attenuated and used as a probe pulse, that was focused along with the THz pulse onto the sample inside the cryostat. In the TKE experiment, the 800 nm probe pulse polarized at 45° relative to the polarization of the THz pulse was transmitted through the sample. The transmitted probe pulse was depolarized by the THz-field-induced anisotropic responses, resulting in transient birefringence. The signal was measured by a pair of balanced photodiodes after a half-wave plate and a Wollaston prism. 

\noindent \textit{Time-domain THz spectroscopy}\\
\red{In time-domain THz spectroscopy experiments, The THz field was attenuated by a pair of wire-grid polarizers so that the measured signals were in the linear response regime. The transmitted THz waveform was focused into a ZnTe crystal and was overlapped with a gate pulse at 800 nm for the electro-optic sampling. We determined the frequency-dependent complex transmission coefficient by comparing the THz electric field through the sample to that through a reference aperture of the same size. From the measured complex transmission coefficient, we numerically extracted the real and imaginary part of the dielectric permittivity as a function of frequency.}

\noindent \textit{MD simulation}\\
The Raman spectrum was calculated via an MD approach as described before\cite{thomas2013computing,sharma2020elucidating,sharma2020lattice}:
\begin{equation}
I_{ij}(\omega)=\frac{\omega}{1-exp(-\frac{h\omega}{K_{b}T})}\int<\alpha_{ij}(\tau)\alpha_{ij}(t+\tau)>_{\tau}e^{-iwt}dt
\end{equation}
where $I_{ij}(\omega)$ is the Raman scattering intensity at frequency $\omega$, and $\alpha_{ij}$ is the electronic polarizability tensor obtained from DFPT\cite{Gonze1995Aia,baroni1987green,baroni2001phonons}. The polarizability $\alpha_{ij}(t)$ was calculated based on snapshots of MD trajectories. We used Wiener-Khinchin theorem to calculate the auto-correlation function of the Raman susceptibility.\cite{wiener1930generalized,khintchine1934korrelationstheorie} The MD simulation is first equilibriated for 10 ps, and then sampled every 100 fs with a total 60 ps time window. The MD and Raman simulation were conducted using Quantum-Espresso\cite{giannozzi2009quantum,giannozzi2017advanced}  and the temperature was controlled with the Nos\'e-Hoover thermostat (see Supplementary Note 3).\cite{nose1984unified,hoover1985canonical}

\noindent\textbf{Data availability}
All data that support the findings of this study are available from the corresponding authors on reasonable request.

\noindent\textbf{Acknowledgments} 
Z.Z., Z.-J.L. and K.A.N acknowledge support from the U.S. Department of Energy, Office of Basic Energy Sciences, under Award No. DE-SC0019126. E.B. acknowledge support from the Robert A. Welch Foundation (grant F-2092-20220331). Y.-C.C. acknowledges direct funding from the MIT UROP. J. Z. and A. M. R. acknowledge the support from U.S. Department of Energy, Office of Science, Basic Energy Sciences, under Award No. DE-FG02-07ER46431. National Energy Research Scientific Computing Center (NERSC) provides the computational support by Office of Science User Facility located at Lawrence Berkeley National Laboratory, operated under Contract No. DE-AC02-05CH11231. N.S.D., W.P., N.B., and W.A.T. acknowledge support from the U.S. Department of Energy, Office of Science, Basic Energy Sciences under award DE-SC0019345.

\noindent\textbf{Funding}
U.S. Department of Energy, Office of Basic Energy Sciences, Award No. DE-SC0019126 (ZZ, ZJL, KAN)
Robert A. Welch Foundation grant F-2092-20220331(EB)
U.S. Department of Energy, Office of Science, Basic Energy Sciences, Award No. DE-FG02-07ER46431 (JZ, AMR)
Office of Science User Facility located at Lawrence Berkeley National Laboratory No. DE-AC02-05CH11231 (JZ, AMR)
U.S. Department of Energy, Office of Science, Basic Energy Sciences, Award No. DE-SC0019345. (NSD, WP, NB, WAT)
MIT UROP (YCC)

\noindent\textbf{Author contributions} Z.Z. and E.B. designed the project. Z.Z. and Z.-J.L. performed the THz measurements, assisted by Y.-C.C.. N.S.D. performed the steady-state Raman measurements. W.P. and N.B. synthesized the 2DHP crystals. J.Z. performed the MD and Raman simulations, supported by Z.D.. Z.Z., J.Z., and Z.-J.L. analyzed the data. Z.Z., J.Z., Z.-J.L., E.B. and K.A.N. wrote the paper with inputs from all authors. E.B., W.A.T., A.M.R. and K.A.N. supervised the research.

\noindent\textbf{Competing interests} The authors declare no competing interests.

\newpage
\section*{References}

\footnotesize
\bibliographystyle{naturemag}
\bibliography{paper}

\begin{thebibliography}{10}
\expandafter\ifx\csname url\endcsname\relax
  \def\url#1{\texttt{#1}}\fi
\expandafter\ifx\csname urlprefix\endcsname\relax\def\urlprefix{URL }\fi
\providecommand{\bibinfo}[2]{#2}
\providecommand{\eprint}[2][]{\url{#2}}

\bibitem{gong2018electron}
\bibinfo{author}{Gong, X.} \emph{et~al.}
\newblock \bibinfo{title}{Electron--phonon interaction in efficient perovskite
  blue emitters}.
\newblock \emph{\bibinfo{journal}{Nature materials}}
  \textbf{\bibinfo{volume}{17}}, \bibinfo{pages}{550--556}
  (\bibinfo{year}{2018}).

\bibitem{grancini2019dimensional}
\bibinfo{author}{Grancini, G.} \& \bibinfo{author}{Nazeeruddin, M.~K.}
\newblock \bibinfo{title}{Dimensional tailoring of hybrid perovskites for
  photovoltaics}.
\newblock \emph{\bibinfo{journal}{Nature Reviews Materials}}
  \textbf{\bibinfo{volume}{4}}, \bibinfo{pages}{4--22} (\bibinfo{year}{2019}).

\bibitem{smith2019tuning}
\bibinfo{author}{Smith, M.~D.}, \bibinfo{author}{Connor, B.~A.} \&
  \bibinfo{author}{Karunadasa, H.~I.}
\newblock \bibinfo{title}{Tuning the luminescence of layered halide
  perovskites}.
\newblock \emph{\bibinfo{journal}{Chemical Reviews}}
  \textbf{\bibinfo{volume}{119}}, \bibinfo{pages}{3104--3139}
  (\bibinfo{year}{2019}).

\bibitem{mauck2019excitons}
\bibinfo{author}{Mauck, C.~M.} \& \bibinfo{author}{Tisdale, W.~A.}
\newblock \bibinfo{title}{Excitons in 2d organic--inorganic halide
  perovskites}.
\newblock \emph{\bibinfo{journal}{Trends in Chemistry}}
  \textbf{\bibinfo{volume}{1}}, \bibinfo{pages}{380--393}
  (\bibinfo{year}{2019}).

\bibitem{li20212d}
\bibinfo{author}{Li, X.}, \bibinfo{author}{Hoffman, J.~M.} \&
  \bibinfo{author}{Kanatzidis, M.~G.}
\newblock \bibinfo{title}{The {2D} halide perovskite rulebook: how the spacer
  influences everything from the structure to optoelectronic device
  efficiency}.
\newblock \emph{\bibinfo{journal}{Chemical Reviews}}
  \textbf{\bibinfo{volume}{121}}, \bibinfo{pages}{2230--2291}
  (\bibinfo{year}{2021}).

\bibitem{shi2018two}
\bibinfo{author}{Shi, E.} \emph{et~al.}
\newblock \bibinfo{title}{Two-dimensional halide perovskite nanomaterials and
  heterostructures}.
\newblock \emph{\bibinfo{journal}{Chemical Society Reviews}}
  \textbf{\bibinfo{volume}{47}}, \bibinfo{pages}{6046--6072}
  (\bibinfo{year}{2018}).

\bibitem{saparov2016organic}
\bibinfo{author}{Saparov, B.} \& \bibinfo{author}{Mitzi, D.~B.}
\newblock \bibinfo{title}{Organic--inorganic perovskites: structural
  versatility for functional materials design}.
\newblock \emph{\bibinfo{journal}{Chemical Reviews}}
  \textbf{\bibinfo{volume}{116}}, \bibinfo{pages}{4558--4596}
  (\bibinfo{year}{2016}).

\bibitem{paritmongkol2019synthetic}
\bibinfo{author}{Paritmongkol, W.} \emph{et~al.}
\newblock \bibinfo{title}{Synthetic variation and structural trends in layered
  two-dimensional alkylammonium lead halide perovskites}.
\newblock \emph{\bibinfo{journal}{Chemistry of Materials}}
  \textbf{\bibinfo{volume}{31}}, \bibinfo{pages}{5592--5607}
  (\bibinfo{year}{2019}).

\bibitem{hang2011metal}
\bibinfo{author}{Hang, T.}, \bibinfo{author}{Zhang, W.}, \bibinfo{author}{Ye,
  H.-Y.} \& \bibinfo{author}{Xiong, R.-G.}
\newblock \bibinfo{title}{Metal-organic complex ferroelectrics}.
\newblock \emph{\bibinfo{journal}{Chemical Society Reviews}}
  \textbf{\bibinfo{volume}{40}}, \bibinfo{pages}{3577--3598}
  (\bibinfo{year}{2011}).

\bibitem{chen2019two}
\bibinfo{author}{Chen, X.-G.} \emph{et~al.}
\newblock \bibinfo{title}{Two-dimensional layered perovskite ferroelectric with
  giant piezoelectric voltage coefficient}.
\newblock \emph{\bibinfo{journal}{Journal of the American Chemical Society}}
  \textbf{\bibinfo{volume}{142}}, \bibinfo{pages}{1077--1082}
  (\bibinfo{year}{2019}).

\bibitem{kim2021chiral}
\bibinfo{author}{Kim, Y.-H.} \emph{et~al.}
\newblock \bibinfo{title}{Chiral-induced spin selectivity enables a
  room-temperature spin light-emitting diode}.
\newblock \emph{\bibinfo{journal}{Science}} \textbf{\bibinfo{volume}{371}},
  \bibinfo{pages}{1129--1133} (\bibinfo{year}{2021}).

\bibitem{li2017chemically}
\bibinfo{author}{Li, W.} \emph{et~al.}
\newblock \bibinfo{title}{Chemically diverse and multifunctional hybrid
  organic--inorganic perovskites}.
\newblock \emph{\bibinfo{journal}{Nature Reviews Materials}}
  \textbf{\bibinfo{volume}{2}}, \bibinfo{pages}{1--18} (\bibinfo{year}{2017}).

\bibitem{menahem2021strongly}
\bibinfo{author}{Menahem, M.} \emph{et~al.}
\newblock \bibinfo{title}{Strongly anharmonic octahedral tilting in
  two-dimensional hybrid halide perovskites}.
\newblock \emph{\bibinfo{journal}{ACS Nano}} \textbf{\bibinfo{volume}{15}},
  \bibinfo{pages}{10153--10162} (\bibinfo{year}{2021}).

\bibitem{mayers2018lattice}
\bibinfo{author}{Mayers, M.~Z.}, \bibinfo{author}{Tan, L.~Z.},
  \bibinfo{author}{Egger, D.~A.}, \bibinfo{author}{Rappe, A.~M.} \&
  \bibinfo{author}{Reichman, D.~R.}
\newblock \bibinfo{title}{How lattice and charge fluctuations control carrier
  dynamics in halide perovskites}.
\newblock \emph{\bibinfo{journal}{Nano Letters}} \textbf{\bibinfo{volume}{18}},
  \bibinfo{pages}{8041--8046} (\bibinfo{year}{2018}).

\bibitem{egger2018remains}
\bibinfo{author}{Egger, D.~A.} \emph{et~al.}
\newblock \bibinfo{title}{What remains unexplained about the properties of
  halide perovskites?}
\newblock \emph{\bibinfo{journal}{Advanced Materials}}
  \textbf{\bibinfo{volume}{30}}, \bibinfo{pages}{1800691}
  (\bibinfo{year}{2018}).

\bibitem{miyata2018ferroelectric}
\bibinfo{author}{Miyata, K.} \& \bibinfo{author}{Zhu, X.-Y.}
\newblock \bibinfo{title}{Ferroelectric large polarons}.
\newblock \emph{\bibinfo{journal}{Nature Materials}}
  \textbf{\bibinfo{volume}{17}}, \bibinfo{pages}{379--381}
  (\bibinfo{year}{2018}).

\bibitem{rivett2018long}
\bibinfo{author}{Rivett, J.~P.} \emph{et~al.}
\newblock \bibinfo{title}{Long-lived polarization memory in the electronic
  states of lead-halide perovskites from local structural dynamics}.
\newblock \emph{\bibinfo{journal}{Nature Communications}}
  \textbf{\bibinfo{volume}{9}}, \bibinfo{pages}{1--8} (\bibinfo{year}{2018}).

\bibitem{dhanabalan2020directional}
\bibinfo{author}{Dhanabalan, B.} \emph{et~al.}
\newblock \bibinfo{title}{Directional anisotropy of the vibrational modes in
  {2D}-layered perovskites}.
\newblock \emph{\bibinfo{journal}{ACS Nano}} \textbf{\bibinfo{volume}{14}},
  \bibinfo{pages}{4689--4697} (\bibinfo{year}{2020}).

\bibitem{hlinka2008coexistence}
\bibinfo{author}{Hlinka, J.} \emph{et~al.}
\newblock \bibinfo{title}{Coexistence of the phonon and relaxation soft modes
  in the terahertz dielectric response of tetragonal {BaTiO}$_3$}.
\newblock \emph{\bibinfo{journal}{Physical Review Letters}}
  \textbf{\bibinfo{volume}{101}}, \bibinfo{pages}{167402}
  (\bibinfo{year}{2008}).

\bibitem{yaffe2017local}
\bibinfo{author}{Yaffe, O.} \emph{et~al.}
\newblock \bibinfo{title}{Local polar fluctuations in lead halide perovskite
  crystals}.
\newblock \emph{\bibinfo{journal}{Physical Review Letters}}
  \textbf{\bibinfo{volume}{118}}, \bibinfo{pages}{136001}
  (\bibinfo{year}{2017}).

\bibitem{dahod2020low}
\bibinfo{author}{Dahod, N.~S.}, \bibinfo{author}{France-Lanord, A.},
  \bibinfo{author}{Paritmongkol, W.}, \bibinfo{author}{Grossman, J.~C.} \&
  \bibinfo{author}{Tisdale, W.~A.}
\newblock \bibinfo{title}{Low-frequency {Raman} spectrum of {2D} layered
  perovskites: Local atomistic motion or superlattice modes?}
\newblock \emph{\bibinfo{journal}{The Journal of Chemical Physics}}
  \textbf{\bibinfo{volume}{153}}, \bibinfo{pages}{044710}
  (\bibinfo{year}{2020}).

\bibitem{yan1987impulsive}
\bibinfo{author}{Yan, Y.-X.} \& \bibinfo{author}{Nelson, K.~A.}
\newblock \bibinfo{title}{Impulsive stimulated light scattering. ii. comparison
  to frequency-domain light-scattering spectroscopy}.
\newblock \emph{\bibinfo{journal}{The Journal of Chemical Physics}}
  \textbf{\bibinfo{volume}{87}}, \bibinfo{pages}{6257--6265}
  (\bibinfo{year}{1987}).

\bibitem{dougherty1992femtosecond}
\bibinfo{author}{Dougherty, T.~P.} \emph{et~al.}
\newblock \bibinfo{title}{Femtosecond resolution of soft mode dynamics in
  structural phase transitions}.
\newblock \emph{\bibinfo{journal}{Science}} \textbf{\bibinfo{volume}{258}},
  \bibinfo{pages}{770--774} (\bibinfo{year}{1992}).

\bibitem{hoffmann2009terahertz}
\bibinfo{author}{Hoffmann, M.~C.}, \bibinfo{author}{Brandt, N.~C.},
  \bibinfo{author}{Hwang, H.~Y.}, \bibinfo{author}{Yeh, K.-L.} \&
  \bibinfo{author}{Nelson, K.~A.}
\newblock \bibinfo{title}{Terahertz {Kerr} effect}.
\newblock \emph{\bibinfo{journal}{Applied Physics Letters}}
  \textbf{\bibinfo{volume}{95}}, \bibinfo{pages}{231105}
  (\bibinfo{year}{2009}).

\bibitem{zhong2008optical}
\bibinfo{author}{Zhong, Q.} \& \bibinfo{author}{Fourkas, J.~T.}
\newblock \bibinfo{title}{Optical {Kerr} effect spectroscopy of simple
  liquids}.
\newblock \emph{\bibinfo{journal}{The Journal of Physical Chemistry B}}
  \textbf{\bibinfo{volume}{112}}, \bibinfo{pages}{15529--15539}
  (\bibinfo{year}{2008}).

\bibitem{miyata2017large}
\bibinfo{author}{Miyata, K.} \emph{et~al.}
\newblock \bibinfo{title}{Large polarons in lead halide perovskites}.
\newblock \emph{\bibinfo{journal}{Science Advances}}
  \textbf{\bibinfo{volume}{3}}, \bibinfo{pages}{e1701217}
  (\bibinfo{year}{2017}).

\bibitem{zhu2016screening}
\bibinfo{author}{Zhu, H.} \emph{et~al.}
\newblock \bibinfo{title}{Screening in crystalline liquids protects energetic
  carriers in hybrid perovskites}.
\newblock \emph{\bibinfo{journal}{Science}} \textbf{\bibinfo{volume}{353}},
  \bibinfo{pages}{1409--1413} (\bibinfo{year}{2016}).

\bibitem{maehrlein2021decoding}
\bibinfo{author}{Maehrlein, S.~F.} \emph{et~al.}
\newblock \bibinfo{title}{Decoding ultrafast polarization responses in lead
  halide perovskites by the two-dimensional optical {Kerr} effect}.
\newblock \emph{\bibinfo{journal}{Proceedings of the National Academy of
  Sciences}} \textbf{\bibinfo{volume}{118}}, \bibinfo{pages}{e2022268118}
  (\bibinfo{year}{2021}).

\bibitem{melnikov2022terahertz}
\bibinfo{author}{Melnikov, A.}, \bibinfo{author}{Anikeeva, V.},
  \bibinfo{author}{Semenova, O.} \& \bibinfo{author}{Chekalin, S.}
\newblock \bibinfo{title}{Terahertz kerr effect in a methylammonium lead
  bromide perovskite crystal}.
\newblock \emph{\bibinfo{journal}{Physical Review B}}
  \textbf{\bibinfo{volume}{105}}, \bibinfo{pages}{174304}
  (\bibinfo{year}{2022}).

\bibitem{elgabarty2020energy}
\bibinfo{author}{Elgabarty, H.} \emph{et~al.}
\newblock \bibinfo{title}{Energy transfer within the hydrogen bonding network
  of water following resonant terahertz excitation}.
\newblock \emph{\bibinfo{journal}{Science Advances}}
  \textbf{\bibinfo{volume}{6}}, \bibinfo{pages}{eaay7074}
  (\bibinfo{year}{2020}).

\bibitem{zhao2020ultrafast}
\bibinfo{author}{Zhao, H.} \emph{et~al.}
\newblock \bibinfo{title}{Ultrafast hydrogen bond dynamics of liquid water
  revealed by terahertz-induced transient birefringence}.
\newblock \emph{\bibinfo{journal}{Light: Science \& Applications}}
  \textbf{\bibinfo{volume}{9}}, \bibinfo{pages}{1--10} (\bibinfo{year}{2020}).

\bibitem{reyes2020unraveling}
\bibinfo{author}{Reyes-Martinez, M.~A.} \emph{et~al.}
\newblock \bibinfo{title}{Unraveling the elastic properties of (quasi)
  two-dimensional hybrid perovskites: a joint experimental and theoretical
  study}.
\newblock \emph{\bibinfo{journal}{ACS Applied Materials \& Interfaces}}
  \textbf{\bibinfo{volume}{12}}, \bibinfo{pages}{17881--17892}
  (\bibinfo{year}{2020}).

\bibitem{khalsa2021ultrafast}
\bibinfo{author}{Khalsa, G.}, \bibinfo{author}{Benedek, N.~A.} \&
  \bibinfo{author}{Moses, J.}
\newblock \bibinfo{title}{Ultrafast control of material optical properties via
  the infrared resonant {Raman} effect}.
\newblock \emph{\bibinfo{journal}{Physical Review X}}
  \textbf{\bibinfo{volume}{11}}, \bibinfo{pages}{021067}
  (\bibinfo{year}{2021}).

\bibitem{forst_nonlinear_2011}
\bibinfo{author}{F{\"o}rst, M.} \emph{et~al.}
\newblock \bibinfo{title}{Nonlinear phononics as an ultrafast route to lattice
  control}.
\newblock \emph{\bibinfo{journal}{Nature Physics}}
  \textbf{\bibinfo{volume}{7}}, \bibinfo{pages}{854--856}
  (\bibinfo{year}{2011}).

\bibitem{juraschek2018sum}
\bibinfo{author}{Juraschek, D.~M.} \& \bibinfo{author}{Maehrlein, S.~F.}
\newblock \bibinfo{title}{Sum-frequency ionic {Raman} scattering}.
\newblock \emph{\bibinfo{journal}{Physical Review B}}
  \textbf{\bibinfo{volume}{97}}, \bibinfo{pages}{174302}
  (\bibinfo{year}{2018}).

\bibitem{aku2005long}
\bibinfo{author}{Aku-Leh, C.}, \bibinfo{author}{Zhao, J.},
  \bibinfo{author}{Merlin, R.}, \bibinfo{author}{Menendez, J.} \&
  \bibinfo{author}{Cardona, M.}
\newblock \bibinfo{title}{Long-lived optical phonons in {ZnO} studied with
  impulsive stimulated {Raman} scattering}.
\newblock \emph{\bibinfo{journal}{Physical Review B}}
  \textbf{\bibinfo{volume}{71}}, \bibinfo{pages}{205211}
  (\bibinfo{year}{2005}).

\bibitem{johnson2019distinguishing}
\bibinfo{author}{Johnson, C.~L.}, \bibinfo{author}{Knighton, B.~E.} \&
  \bibinfo{author}{Johnson, J.~A.}
\newblock \bibinfo{title}{Distinguishing nonlinear terahertz excitation
  pathways with two-dimensional spectroscopy}.
\newblock \emph{\bibinfo{journal}{Physical Review Letters}}
  \textbf{\bibinfo{volume}{122}}, \bibinfo{pages}{073901}
  (\bibinfo{year}{2019}).

\bibitem{maehrlein2017terahertz}
\bibinfo{author}{Maehrlein, S.}, \bibinfo{author}{Paarmann, A.},
  \bibinfo{author}{Wolf, M.} \& \bibinfo{author}{Kampfrath, T.}
\newblock \bibinfo{title}{Terahertz sum-frequency excitation of a
  {Raman}-active phonon}.
\newblock \emph{\bibinfo{journal}{Physical Review Letters}}
  \textbf{\bibinfo{volume}{119}}, \bibinfo{pages}{127402}
  (\bibinfo{year}{2017}).

\bibitem{yan1985impulsive}
\bibinfo{author}{Yan, Y.-X.}, \bibinfo{author}{Gamble~Jr, E.~B.} \&
  \bibinfo{author}{Nelson, K.~A.}
\newblock \bibinfo{title}{Impulsive stimulated scattering: General importance
  in femtosecond laser pulse interactions with matter, and spectroscopic
  applications}.
\newblock \emph{\bibinfo{journal}{The Journal of Chemical Physics}}
  \textbf{\bibinfo{volume}{83}}, \bibinfo{pages}{5391--5399}
  (\bibinfo{year}{1985}).

\bibitem{maradudin1970ionic}
\bibinfo{author}{Maradudin, A.} \& \bibinfo{author}{Wallis, R.}
\newblock \bibinfo{title}{Ionic raman effect. i. scattering by localized
  vibration modes}.
\newblock \emph{\bibinfo{journal}{Physical Review B}}
  \textbf{\bibinfo{volume}{2}}, \bibinfo{pages}{4294} (\bibinfo{year}{1970}).

\bibitem{wallis1971ionic}
\bibinfo{author}{Wallis, R.} \& \bibinfo{author}{Maradudin, A.}
\newblock \bibinfo{title}{Ionic raman effect. ii. the first-order ionic raman
  effect}.
\newblock \emph{\bibinfo{journal}{Physical Review B}}
  \textbf{\bibinfo{volume}{3}}, \bibinfo{pages}{2063} (\bibinfo{year}{1971}).

\bibitem{humphreys1972ionic}
\bibinfo{author}{Humphreys, L.}
\newblock \bibinfo{title}{Ionic raman effect. iii. first-and second-order ionic
  raman effects}.
\newblock \emph{\bibinfo{journal}{Physical Review B}}
  \textbf{\bibinfo{volume}{6}}, \bibinfo{pages}{3886} (\bibinfo{year}{1972}).

\bibitem{li2019terahertz}
\bibinfo{author}{Li, X.} \emph{et~al.}
\newblock \bibinfo{title}{Terahertz field--induced ferroelectricity in quantum
  paraelectric {SrTiO}$_3$}.
\newblock \emph{\bibinfo{journal}{Science}} \textbf{\bibinfo{volume}{364}},
  \bibinfo{pages}{1079--1082} (\bibinfo{year}{2019}).

\bibitem{von2022amplification}
\bibinfo{author}{von Hoegen, A.} \emph{et~al.}
\newblock \bibinfo{title}{Amplification of superconducting fluctuations in
  driven {YBa}$_2${Cu}$_3${O}$_{6+x}$}.
\newblock \emph{\bibinfo{journal}{Physical Review X}}
  \textbf{\bibinfo{volume}{12}}, \bibinfo{pages}{031008}
  (\bibinfo{year}{2022}).

\bibitem{mead2020sum}
\bibinfo{author}{Mead, G.}, \bibinfo{author}{Lin, H.-W.},
  \bibinfo{author}{Magdau, I.-B.}, \bibinfo{author}{Miller~III, T.~F.} \&
  \bibinfo{author}{Blake, G.~A.}
\newblock \bibinfo{title}{Sum-frequency signals in 2d-terahertz-terahertz-raman
  spectroscopy}.
\newblock \emph{\bibinfo{journal}{The Journal of Physical Chemistry B}}
  \textbf{\bibinfo{volume}{124}}, \bibinfo{pages}{8904--8908}
  (\bibinfo{year}{2020}).

\bibitem{gonze1995adiabatic}
\bibinfo{author}{Gonze, X.}
\newblock \bibinfo{title}{Adiabatic density-functional perturbation theory}.
\newblock \emph{\bibinfo{journal}{Physical Review A}}
  \textbf{\bibinfo{volume}{52}}, \bibinfo{pages}{1096} (\bibinfo{year}{1995}).

\bibitem{baroni1987green}
\bibinfo{author}{Baroni, S.}, \bibinfo{author}{Giannozzi, P.} \&
  \bibinfo{author}{Testa, A.}
\newblock \bibinfo{title}{Green’s-function approach to linear response in
  solids}.
\newblock \emph{\bibinfo{journal}{Physical Review Letters}}
  \textbf{\bibinfo{volume}{58}}, \bibinfo{pages}{1861} (\bibinfo{year}{1987}).

\bibitem{baroni2001phonons}
\bibinfo{author}{Baroni, S.}, \bibinfo{author}{De~Gironcoli, S.},
  \bibinfo{author}{Dal~Corso, A.} \& \bibinfo{author}{Giannozzi, P.}
\newblock \bibinfo{title}{Phonons and related crystal properties from
  density-functional perturbation theory}.
\newblock \emph{\bibinfo{journal}{Reviews of Modern Physics}}
  \textbf{\bibinfo{volume}{73}}, \bibinfo{pages}{515} (\bibinfo{year}{2001}).

\bibitem{disa2021engineering}
\bibinfo{author}{Disa, A.~S.}, \bibinfo{author}{Nova, T.~F.} \&
  \bibinfo{author}{Cavalleri, A.}
\newblock \bibinfo{title}{Engineering crystal structures with light}.
\newblock \emph{\bibinfo{journal}{Nature Physics}}
  \textbf{\bibinfo{volume}{17}}, \bibinfo{pages}{1087--1092}
  (\bibinfo{year}{2021}).

\bibitem{grinblat2019ultrafast}
\bibinfo{author}{Grinblat, G.} \emph{et~al.}
\newblock \bibinfo{title}{Ultrafast all-optical modulation in {2D} hybrid
  perovskites}.
\newblock \emph{\bibinfo{journal}{ACS Nano}} \textbf{\bibinfo{volume}{13}},
  \bibinfo{pages}{9504--9510} (\bibinfo{year}{2019}).

\bibitem{sun2016optical}
\bibinfo{author}{Sun, Z.}, \bibinfo{author}{Martinez, A.} \&
  \bibinfo{author}{Wang, F.}
\newblock \bibinfo{title}{Optical modulators with {2D} layered materials}.
\newblock \emph{\bibinfo{journal}{Nature Photonics}}
  \textbf{\bibinfo{volume}{10}}, \bibinfo{pages}{227--238}
  (\bibinfo{year}{2016}).

\bibitem{yeh2008generation}
\bibinfo{author}{Yeh, K.-L.}, \bibinfo{author}{Hebling, J.},
  \bibinfo{author}{Hoffmann, M.~C.} \& \bibinfo{author}{Nelson, K.~A.}
\newblock \bibinfo{title}{Generation of high average power 1 {kHz} shaped thz
  pulses via optical rectification}.
\newblock \emph{\bibinfo{journal}{Optics Communications}}
  \textbf{\bibinfo{volume}{281}}, \bibinfo{pages}{3567--3570}
  (\bibinfo{year}{2008}).

\bibitem{thomas2013computing}
\bibinfo{author}{Thomas, M.}, \bibinfo{author}{Brehm, M.},
  \bibinfo{author}{Fligg, R.}, \bibinfo{author}{V{\"o}hringer, P.} \&
  \bibinfo{author}{Kirchner, B.}
\newblock \bibinfo{title}{Computing vibrational spectra from ab initio
  molecular dynamics}.
\newblock \emph{\bibinfo{journal}{Physical Chemistry Chemical Physics}}
  \textbf{\bibinfo{volume}{15}}, \bibinfo{pages}{6608--6622}
  (\bibinfo{year}{2013}).

\bibitem{sharma2020elucidating}
\bibinfo{author}{Sharma, R.} \emph{et~al.}
\newblock \bibinfo{title}{Elucidating the atomistic origin of anharmonicity in
  tetragonal {CH}$_3${NH}$_3${PbI}$_3$ with {Raman} scattering}.
\newblock \emph{\bibinfo{journal}{Physical Review Materials}}
  \textbf{\bibinfo{volume}{4}}, \bibinfo{pages}{092401} (\bibinfo{year}{2020}).

\bibitem{sharma2020lattice}
\bibinfo{author}{Sharma, R.} \emph{et~al.}
\newblock \bibinfo{title}{Lattice mode symmetry analysis of the orthorhombic
  phase of methylammonium lead iodide using polarized {Raman}}.
\newblock \emph{\bibinfo{journal}{Physical Review Materials}}
  \textbf{\bibinfo{volume}{4}}, \bibinfo{pages}{051601} (\bibinfo{year}{2020}).

\bibitem{Gonze1995Aia}
\bibinfo{author}{Gonze, X.}
\newblock \bibinfo{title}{Adiabatic density-functional perturbation theory}.
\newblock \emph{\bibinfo{journal}{Phys. Rev. A}} \textbf{\bibinfo{volume}{52}},
  \bibinfo{pages}{1096--1114} (\bibinfo{year}{1995}).

\bibitem{wiener1930generalized}
\bibinfo{author}{Wiener, N.}
\newblock \bibinfo{title}{Generalized harmonic analysis}.
\newblock \emph{\bibinfo{journal}{Acta mathematica}}
  \textbf{\bibinfo{volume}{55}}, \bibinfo{pages}{117--258}
  (\bibinfo{year}{1930}).

\bibitem{khintchine1934korrelationstheorie}
\bibinfo{author}{Khintchine, A.}
\newblock \bibinfo{title}{Korrelationstheorie der station{\"a}ren
  stochastischen prozesse}.
\newblock \emph{\bibinfo{journal}{Mathematische Annalen}}
  \textbf{\bibinfo{volume}{109}}, \bibinfo{pages}{604--615}
  (\bibinfo{year}{1934}).

\bibitem{giannozzi2009quantum}
\bibinfo{author}{Giannozzi, P.} \emph{et~al.}
\newblock \bibinfo{title}{{QUANTUM ESPRESSO}: a modular and open-source
  software project for quantum simulations of materials}.
\newblock \emph{\bibinfo{journal}{Journal of Physics: Condensed Matter}}
  \textbf{\bibinfo{volume}{21}}, \bibinfo{pages}{395502}
  (\bibinfo{year}{2009}).

\bibitem{giannozzi2017advanced}
\bibinfo{author}{Giannozzi, P.} \emph{et~al.}
\newblock \bibinfo{title}{Advanced capabilities for materials modelling with
  {Quantum ESPRESSO}}.
\newblock \emph{\bibinfo{journal}{Journal of Physics: Condensed Matter}}
  \textbf{\bibinfo{volume}{29}}, \bibinfo{pages}{465901}
  (\bibinfo{year}{2017}).

\bibitem{nose1984unified}
\bibinfo{author}{Nos{\'e}, S.}
\newblock \bibinfo{title}{A unified formulation of the constant temperature
  molecular dynamics methods}.
\newblock \emph{\bibinfo{journal}{The Journal of Chemical Physics}}
  \textbf{\bibinfo{volume}{81}}, \bibinfo{pages}{511--519}
  (\bibinfo{year}{1984}).

\bibitem{hoover1985canonical}
\bibinfo{author}{Hoover, W.~G.}
\newblock \bibinfo{title}{Canonical dynamics: Equilibrium phase-space
  distributions}.
\newblock \emph{\bibinfo{journal}{Physical Review A}}
  \textbf{\bibinfo{volume}{31}}, \bibinfo{pages}{1695} (\bibinfo{year}{1985}).

\bibitem{hase1998dynamics}
\bibinfo{author}{Hase, M.}, \bibinfo{author}{Mizoguchi, K.},
  \bibinfo{author}{Harima, H.}, \bibinfo{author}{Nakashima, S.-i.} \&
  \bibinfo{author}{Sakai, K.}
\newblock \bibinfo{title}{Dynamics of coherent phonons in bismuth generated by
  ultrashort laser pulses}.
\newblock \emph{\bibinfo{journal}{Physical Review B}}
  \textbf{\bibinfo{volume}{58}}, \bibinfo{pages}{5448} (\bibinfo{year}{1998}).

\end{thebibliography}


\begin{thebibliography}{}
\expandafter\ifx\csname url\endcsname\relax
  \def\url#1{\texttt{#1}}\fi
\expandafter\ifx\csname urlprefix\endcsname\relax\def\urlprefix{URL }\fi
\providecommand{\bibinfo}[2]{#2}
\providecommand{\eprint}[2][]{\url{#2}}

\end{thebibliography}
\newpage
\normalsize

\FloatBarrier

\begin{figure}
	\centering
	\includegraphics[width=0.95\linewidth]{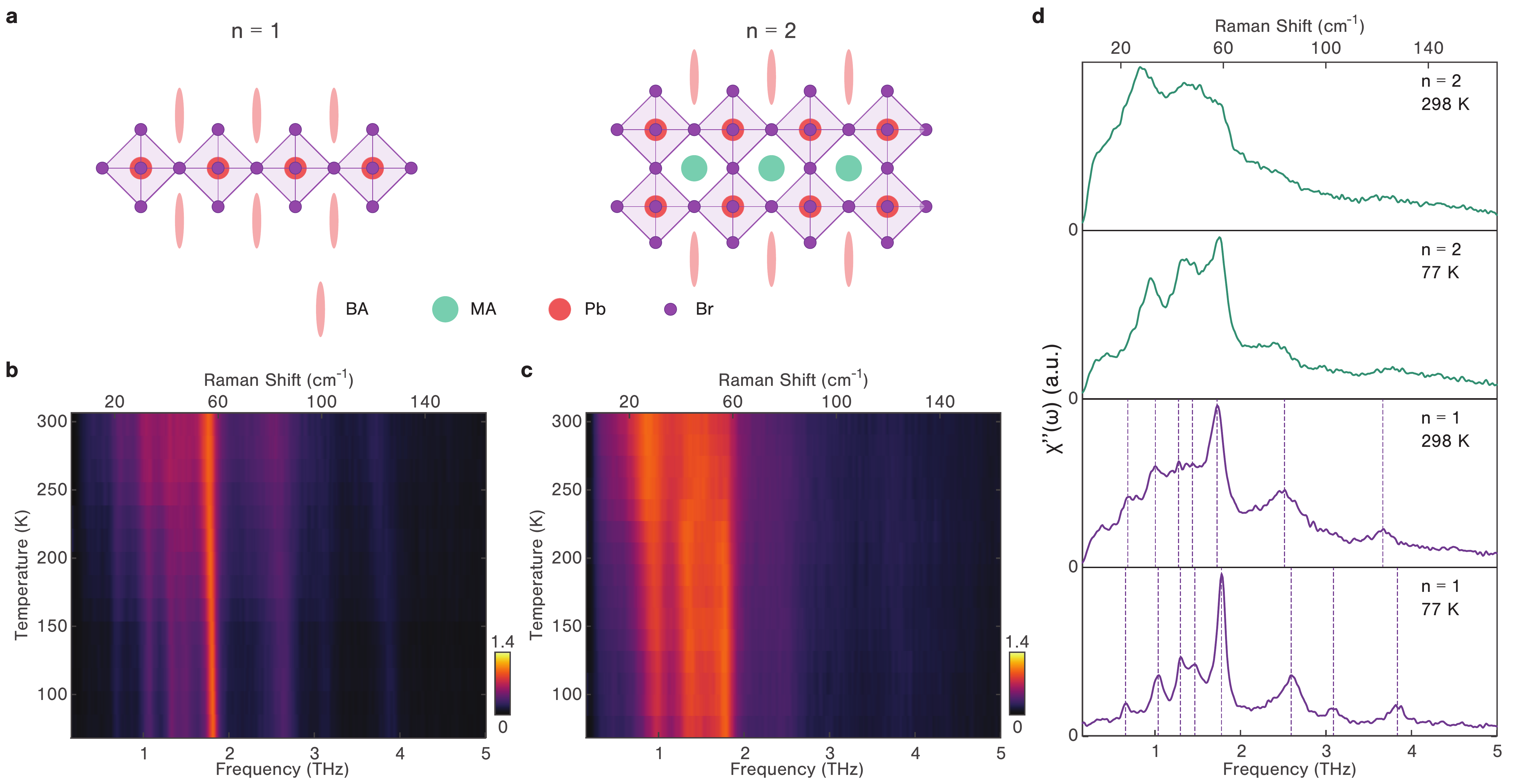} 
	\caption{\label{fig:Fig1}
	\textbf{Crystal structure and static Raman responses of 2DHPs. a,} Schematic illustration of crystal structure for both single-layer (n=1) and double-layer (n=2) bromide perovskites. BA: butylammonium; MA: methylammonium; Pb: lead; Br: bromide. \textbf{b,} and \textbf{c,} Temperature-dependent Raman spectra of n=1 and n=2 2DHPs from 77 to 298 K. \textbf{d,} Selected Raman spectra of n=1 (bottom) and n=2 (top) 2DHPs at 77 K (bright purple and green) and 298 K (light purple and green). All the phonon modes are indicated by dashed lines. The quasi-elastic peaks are marked by black arrows.}
\end{figure}

\begin{figure}
	\centering
	\includegraphics[width=0.75\linewidth]{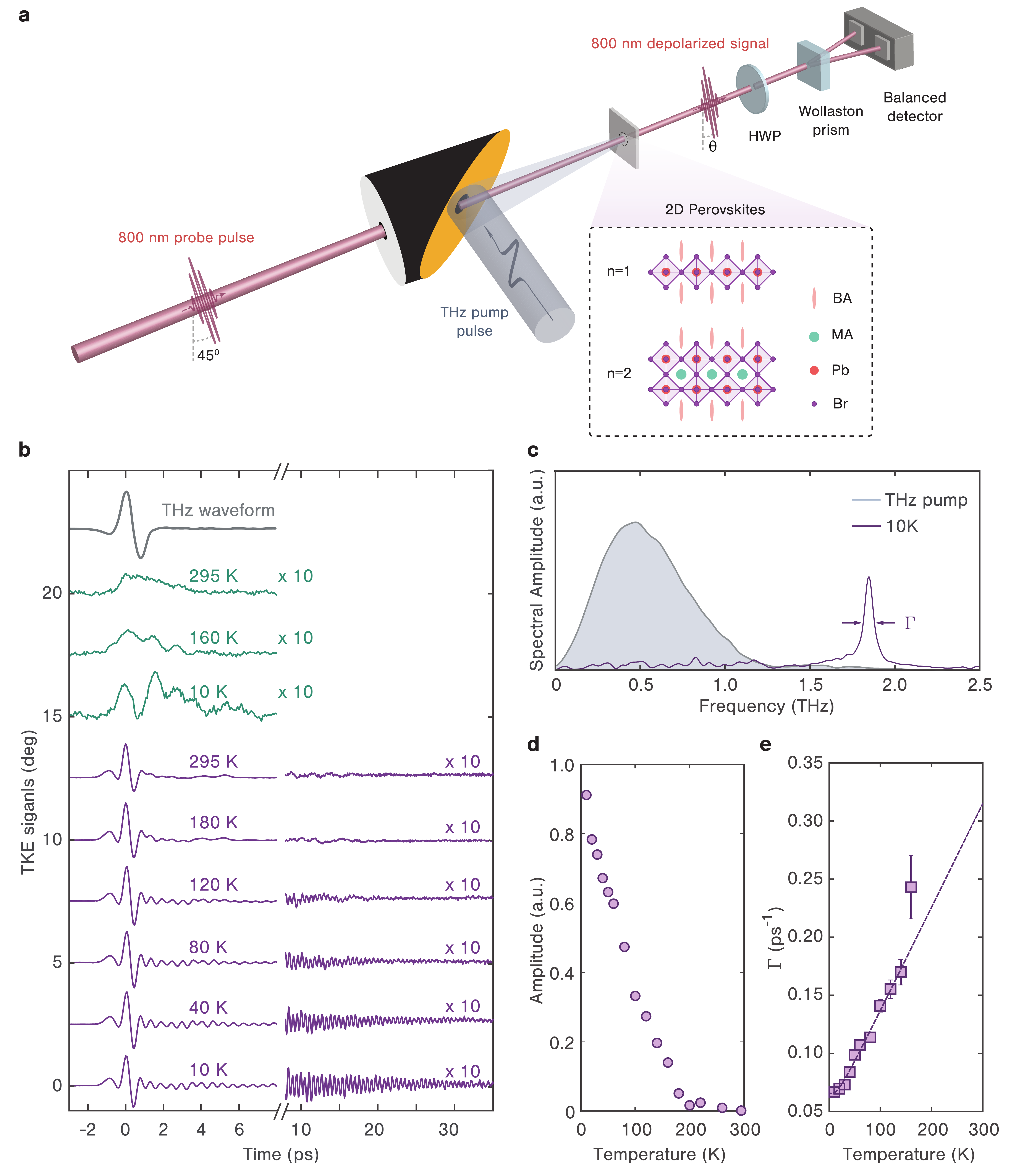}
	\caption{\label{fig:Fig2}  
	\textbf{THz Kerr effect spectroscopy measurements. a,} The single-cycle THz pump is focused on both 2DHP single crystals to induce a nonlinear polarization response. The time-delayed 800-nm probe pulse is polarized at 45$^{\circ}$ relative to the vertical THz polarization and the transiently depolarized signal is measured by a balanced detection scheme. HWP, half-wave plate. \textbf{b,} Time-resolved THz-Raman signals for both n=1 (purple) and n=2 (green) 2DHPs at various temperatures. The TKE signals for n=2 2DHP and the long-lived oscillations in the n=1 sample are magnified by 10. Data are vertically shifted for clarity. The THz pump waveform (grey) is also shown at the top. \textbf{c,} Fourier transform analysis of the oscillatory signal in n=1 sample at 10 K in a reveals a single peak at 1.8 THz, which is above the spectral component of the incident THz pulse (grey area). \textbf{d,} Temperature dependence of the mode amplitude. The amplitude becomes non-zero below 200 K and increases monotonically as the temperature is decreased.  \textbf{e,} The mode dephasing rate as a function of temperature below 200 K. The dashed blue curve is a fit to an anharmonic decay model\cite{hase1998dynamics}. The error bars represent the 95$\%$ confidence interval.}
\end{figure}

\begin{figure}
	\centering
	\includegraphics[width=1\linewidth]{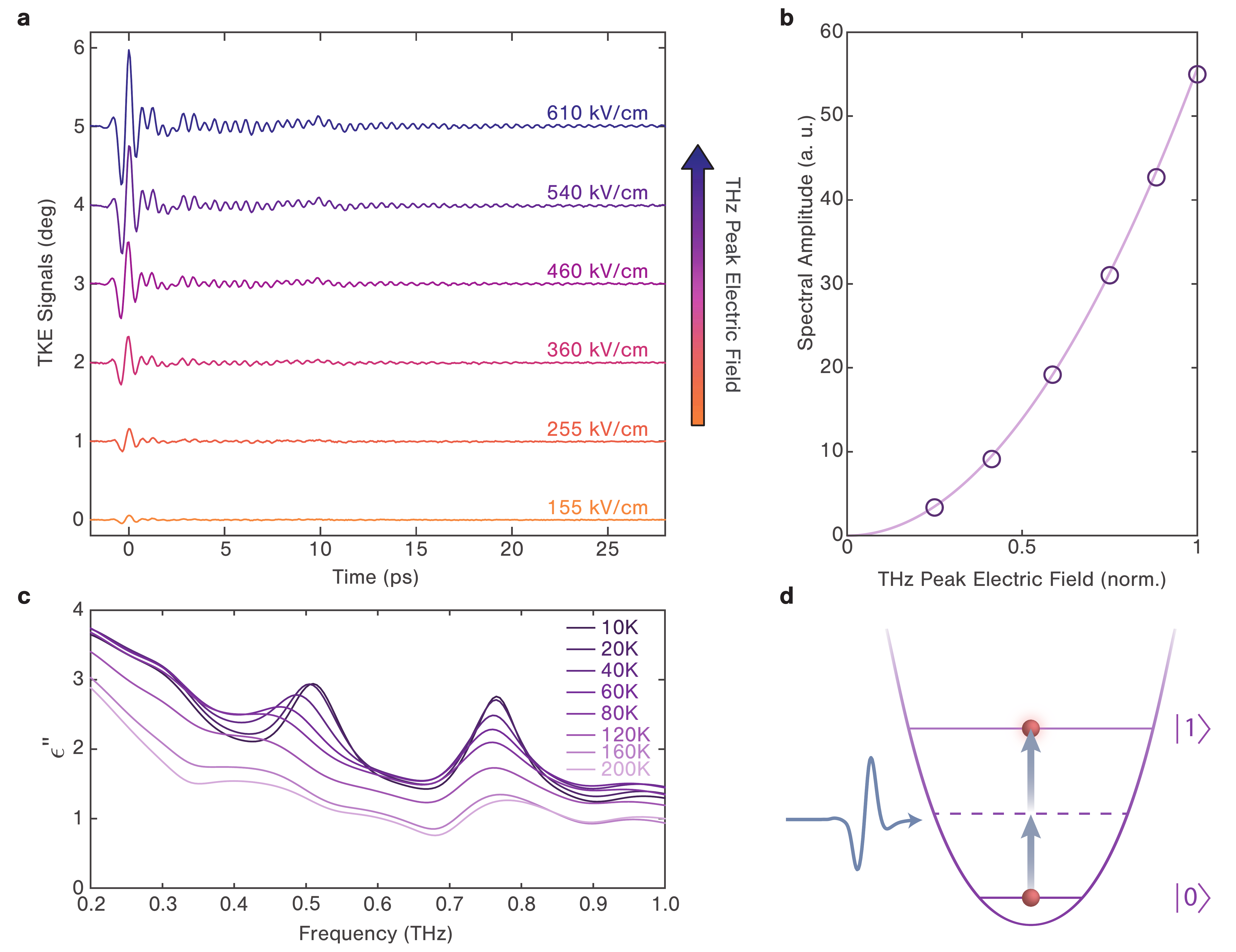}
	\caption{\label{fig:Fig3} 
	\textbf{Generation mechanism of the long-lived phonon oscillation in the n=1 2DHP. a,} TKE signals at 10 K are shown as a function of THz pump electric field strength. Data are vertically shifted for clarity. \textbf{b,} Fourier transform analysis of signals in \textbf{a} shows the spectral amplitudes of the mode as a function of THz pump electric field strength. The light purple line represents a quadratic fit. \textbf{c,} \red{The real part of dielectric permittivity in the n=1 sample as a function of temperature is measured by time-domain THz spectroscopy. The low-temperature curves show two resonance peaks corresponding to two infrared-active phonon modes.}  \textbf{d,} The sum frequency of two incident THz electric-field components is resonant with the transition between the ground and the first excited states to drive the Raman-active mode.}
\end{figure}

\begin{figure}
	\centering
	\includegraphics[width=1\linewidth]{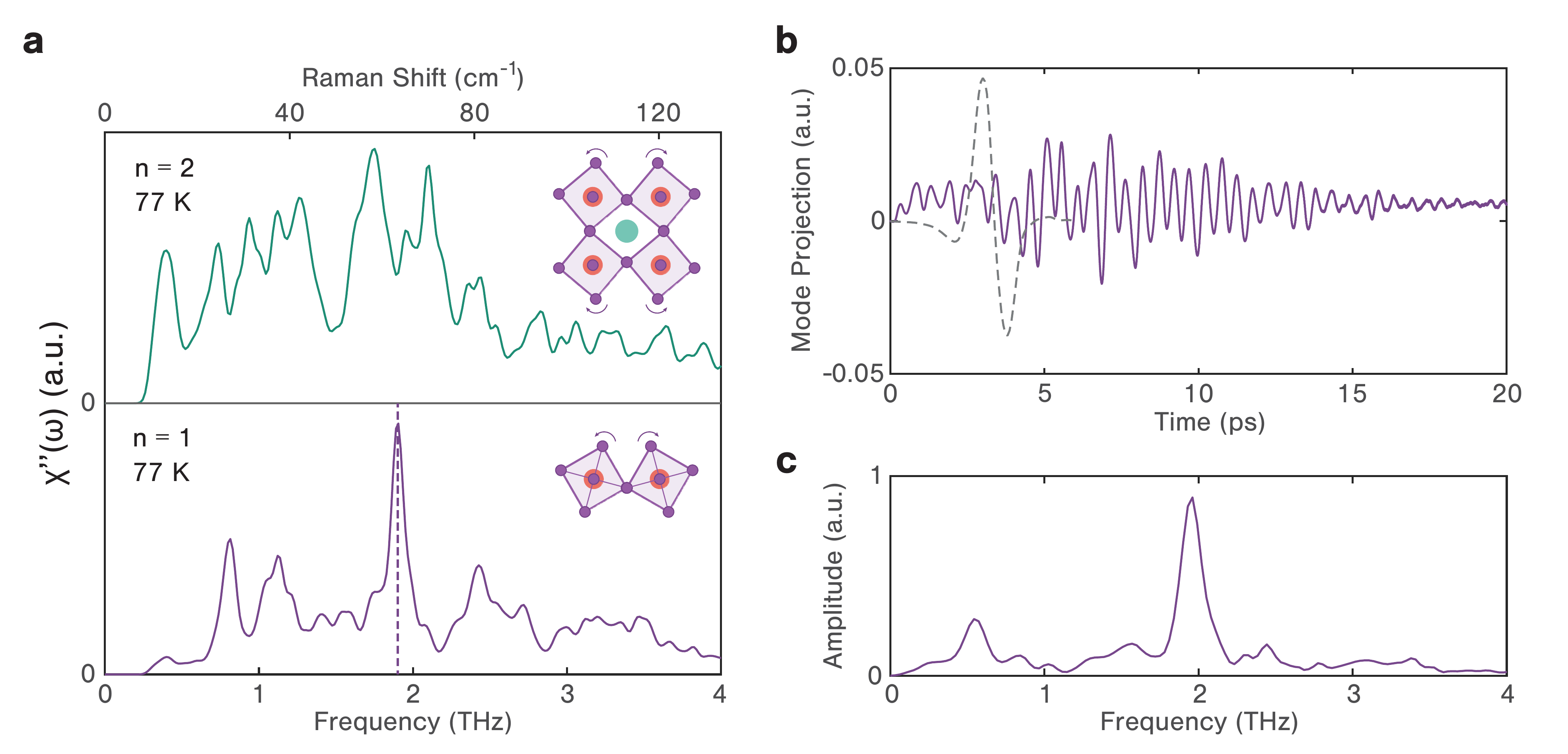}
	\caption{\label{fig:Fig4} 
	\textbf{MD simulation results. a,}  Simulated Raman spectra of both n=1 (bottom) and n=2 (top) 2DHPs at finite temperature (77 K). The insets depict the corresponding octahedral tilting and rotation motion for n=1 and n=2 2DHPs. \textbf{b,} MD simulation trajectory projection onto the 1.8 THz mode (solid purple) along with the input THz electric field waveform (dashed grey). \textbf{c,} Fourier transform of the trajectory projection in \textbf{b}.}
\end{figure}
\FloatBarrier



\section*{Supplementary material}

# Supplementary Information for "Discovery of enhanced lattice dynamics in a single-layered hybrid perovskite"

**Supplementary Note 1: Comparison of TKE signals for n=1 and n=2 2DHPs**

**A. Dielectric properties of lead-bromide 2DHPs**

In our TKE experiments, the THz-field-induced anisotropic responses are detected by the depolarization of the probe pulse at 1.55 eV. This photon energy is much lower than the bandgaps of both n=1 and n=2 2DHPs, and therefore their dielectric properties at 1.55 eV are very similar. To clarify this, we show the optical absorption and photoluminescence spectra[1] in Fig. S1. For the n=1 samples, the lowest excitonic resonance lies at around 3.10 eV and this value is shifted to $\sim$ 2.76 eV for n=2. This set of data demonstrate that our probe photon energy is indeed well below the optical gaps and the TKE signals are not affected by the difference in the dielectric properties of both samples.

**B. TKE response of n-butylammonium bromide**

To identify the origins of the initial bipolar responses in the TKE signals from the n=1 sample, we also performed TKE experiments on crystalline n-butylammonium bromide (98%, SIGMA-ALDRICH) at room temperature. As shown in Fig. S2, the TKE signal from n-butylammonium bromide also displays a strong bipolar response, similar to

what is observed in the n=1 2DHP. Thus, we attribute the origin of the observed bipolar response in the n=1 2DHP to the presence of the BA spacers. However, this bipolar response is not observed in the n=2 2DHP, which has an additional lead bromide octahedral layer and more crucially, the methylammonium organic cations. Therefore, it is likely that this more complicated structure of the n=2 system leads to the suppression of the bipolar response.

**Supplementary Note 2: Sum-frequency excitation of the Raman mode**

In this section, we provide a comparison of the two nonlinear excitation pathways to drive the Raman mode. For a two-photon Raman scattering process, we assume a harmonic lattice potential $V(Q_R) = \frac{1}{2}\Omega_R Q_R^2$, and the corresponding Lagrange equation with a classical Raman type harmonic-oscillator can be derived as [2,3]

$$(\frac{\partial^2}{\partial t^2} + \Gamma_R \frac{\partial}{\partial t} + \Omega_R^2)Q_R = E(t)^2 \frac{\partial \chi}{\partial Q_R}, \tag{1}$$

where $Q_R$ is the normal Raman mode coordinate; $\Gamma_R$ is a phenomenological damping term; $\Omega_R$ is the eigenfrequency of the Raman mode; $E(t)$ represents the driving electric field; and $\chi$ is the linear dielectric susceptibility. Therefore, the excitation of the Raman mode is mediated by the Raman tensor $\frac{\partial \chi}{\partial Q_R}$. Since the driving term on the right-hand side of the equation scales as the square of the pump electric field, which can couple to

the Raman mode through either difference- (i.e., $\Omega_1 - \Omega_2 = \Omega_R$) or sum-frequency (i.e., $\Omega_1 + \Omega_2 = \Omega_R$) components of light, this equation of motion describes impulsive stimulated Raman scattering as well as sum-frequency excitation observed here. It is worth noting that the dielectric responses of hybrid perovskites feature large jumps in the THz range as the frequency decreases across several broad transverse optical phonon resonances. Since for both Raman excitation processes, the pump electric field interacts with virtual electronic dipole transitions, THz off-resonance excitation gives rise to colossal nonlinear polarizability response compared to that in the optical range. This can be also viewed in the time domain as a cloud of electrons bound to a nucleus displaces more strongly in response to a slowly varying electric field.

In contrast, the ionic Raman scattering requires an anharmonic lattice potential and its simplest form can be described as $V(Q_R) = \frac{1}{2}\Omega_R Q_R^2 + \frac{1}{2}\Omega_{IR} Q_{IR}^2 + c Q_{IR} Q_R^2$, where $c$ is the anharmonic coupling coefficient. The corresponding equations of motion are[4]

$$(\frac{\partial^2}{\partial t^2} + \Gamma_{IR}\frac{\partial}{\partial t} + \Omega_{IR}^2 + 2cQ_R)Q_{IR} = Z_{IR}E(t), \tag{2}$$

$$(\frac{\partial^2}{\partial t^2} + \Gamma_R\frac{\partial}{\partial t} + \Omega_R^2)Q_R = cQ_{IR}(t)^2, \tag{3}$$

where in the first equation $Z_{IR}$ is the effective charge of the infrared-active phonon mode. In this case, the Raman mode is activated by anharmonic coupling to the directly driven infrared-active mode. For this process to be efficient, there should exist an

infrared-active phonon mode with its eigenfrequency that matches the sum-frequency excitation condition ($\Omega_{IR} = \frac{1}{2}\Omega_R$, i.e., $\sim 0.9$ THz), which is ruled out by the time-domain THz spectroscopy measurement. Therefore, we confirm that the driven Raman mode excited through large polarizability rather than anharmonicity.

**Supplementary Note 3: MD simulation details**

**A. Finite temperature calculations**

We used a Nosé-Hoover thermostat reference to sample the thermodynamics of the system, with the thermal damping time and the targeted temperature set at 0.1 ps[5] and 77 K, respectively. Figure S3 shows the time evolution of the non-zero tensor elements of the dielectric susceptibility in thermal equilibrium along with the temperature fluctuations (See inset plots). The spontaneous Raman scattering intensity was calculated based on the isotropic average condition,[6]

$$I_{//} \propto \frac{(\omega_{in} - \omega_p)^4}{\omega_p} \frac{45 a_p^2 + 4\gamma_p^2}{45} \frac{1}{1 - exp(-\frac{\hbar \omega_p}{k_B T})}, \quad (4)$$

$$I_{\perp} \propto \frac{(\omega_{in} - \omega_p)^4}{\omega_p} \frac{3\gamma_p^2}{45} \frac{1}{1 - exp(-\frac{\hbar \omega_p}{k_B T})}, \quad (5)$$

where $I_{//}$ and $I_{\perp}$ denote the Raman scattering intensity polarized parallel and perpendicular to the incident light polarization; $\omega_{in}$, $\omega_p$ are the frequencies of the incident and scattered light; and $Q_p$ represents the normal mode coordinate; $a_p$ is the isotropic

polarizability, defined as

$$a_p = \frac{1}{3}(\frac{\partial \chi_{xx}}{\partial Q_p} + \frac{\partial \chi_{yy}}{\partial Q_p} + \frac{\partial \chi_{zz}}{\partial Q_p}), \tag{6}$$

and $\gamma_p$ is the anisotropic polarizability, defined as

$$\gamma_p^2 = \frac{1}{2}(\frac{\partial \chi_{xx}}{\partial Q_p} - \frac{\partial \chi_{yy}}{\partial Q_p})^2 + \frac{1}{2}(\frac{\partial \chi_{yy}}{\partial Q_p} - \frac{\partial \chi_{zz}}{\partial Q_p})^2 + \frac{1}{2}(\frac{\partial \chi_{zz}}{\partial Q_p} - \frac{\partial \chi_{xx}}{\partial Q_p})^2 \\ + 3(\frac{\partial \chi_{xy}}{\partial Q_p})^2 + 3(\frac{\partial \chi_{yz}}{\partial Q_p})^2 + 3(\frac{\partial \chi_{xz}}{\partial Q_p})^2. \tag{7}$$

The Raman tensor $\frac{\partial \chi_{ij}}{\partial Q_p}$ was calculated by computing the time-domain auto-correlation function

$$(\frac{\partial \chi_{ij}}{\partial Q_p})^2 \propto \int <\chi_{ij}(\tau)\chi_{ij}(t+\tau)>_\tau e^{-iw_p t} dt. \tag{8}$$

**B. Real-space analysis**

To identify the lattice displacements corresponding to the Raman peaks, we filtered the trajectories of the system $\vec{X}(t)$ with each peak frequency $\omega_R$ and a window $\Delta\omega$ to get the real-space trajectory with the mode frequency equals to $\omega_R$:

$$\vec{X'}(t, \omega_R) = FT^{-1}\{\Theta(\omega - \omega_R + \Delta\omega)\Theta(\omega - \omega_R - \Delta\omega) \, FT\{\vec{X}(t)\}\} \tag{9}$$

where $\Theta$ is the Heaviside step function, and $FT$ denotes the Fourier transform. We attached GIF files for each distinct lattice motion, corresponding to Raman peaks at 0.7 THz, 1.1 THz, 1.5 THz and 1.8 THz frequencies. The 0.7 THz mode coincides the shearing motion of the two adjacent octahedral layers. The 1.1 THz mode represents the breathing motion of the octahedral layers. The 1.5 THz and 1.8 THz modes correspond to the octahedral bending and twisting.

**Supplementary Note 4: Evaluation of Kerr constant and refractive modulation depth**

In this section, we provide an estimate of the Kerr nonlinear coefficient and the refractive modulation depth of the n=1 2DHP under the intense THz fields. When irradiating the material with a peak THz electric field strength of 610 kV/cm at room temperature, we observe a 1.2% deviation from the balanced signals at the arrival of the THz peak. In the balanced detection scheme using a half-wave plate[7], the differential signal is

$$\frac{\Delta I}{I_0} = \frac{I_1 - I_2}{I_1 + I_2} = \frac{1}{2}sin(2\Gamma), \qquad (10)$$

where $I_1$ and $I_2$ are intensities of the two orthogonally polarized beams measured by a pair of identical photodiodes. The phase shift is $\Gamma = 2\pi \Delta n L / \lambda_{800nm}$, with $L$ being the sample thickness (i.e., $\sim 100 \mu$m) and $\Delta n$ being the THz-induced change in refractive

index. For small polarization rotations, $\Delta I/I_0 \sim 2\pi \Delta n L/\lambda_{800nm}$, from which we calculate $\Delta n = 1.528 \times 10^{-5}$, and the Kerr constant $K = \Delta n/(\lambda_{800nm} E_{THz}^2) = 5.132 \times 10^{-15} m \cdot V^{-2}$. The modulation amplitude is calculated as $10 \times log((I_0 + \Delta I)/(I_0 + \Delta I)) = 0.24$ dB, which means the modulation depth is approximately 2.4 dB/mm. This suggests that 2DHPs are promising candidates for achieving high-speed, all-optical, and broadband refractive modulators.

# References

1. Paritmongkol, W. *et al.* Synthetic variation and structural trends in layered two-dimensional alkylammonium lead halide perovskites. *Chemistry of Materials* **31**, 5592–5607 (2019).

2. Maehrlein, S., Paarmann, A., Wolf, M. & Kampfrath, T. Terahertz sum-frequency excitation of a Raman-active phonon. *Physical Review Letters* **119**, 127402 (2017).

3. Dhar, L., Rogers, J. A. & Nelson, K. A. Time-resolved vibrational spectroscopy in the impulsive limit. *Chemical Reviews* **94**, 157–193 (1994).

4. Juraschek, D. M. & Maehrlein, S. F. Sum-frequency ionic Raman scattering. *Physical Review B* **97**, 174302 (2018).

5. Thompson, A. P. *et al.* Lammps-a flexible simulation tool for particle-based materials modeling at the atomic, meso, and continuum scales. *Computer Physics Communications* **271**, 108171 (2022).

6. Thomas, M., Brehm, M., Fligg, R., Vöhringer, P. & Kirchner, B. Computing vibrational spectra from ab initio molecular dynamics. *Physical Chemistry Chemical Physics* **15**, 6608–6622 (2013).

7. Kumar, V. *et al.* Balanced-detection Raman-induced Kerr-effect spectroscopy. *Physical Review A* **86**, 053810 (2012).

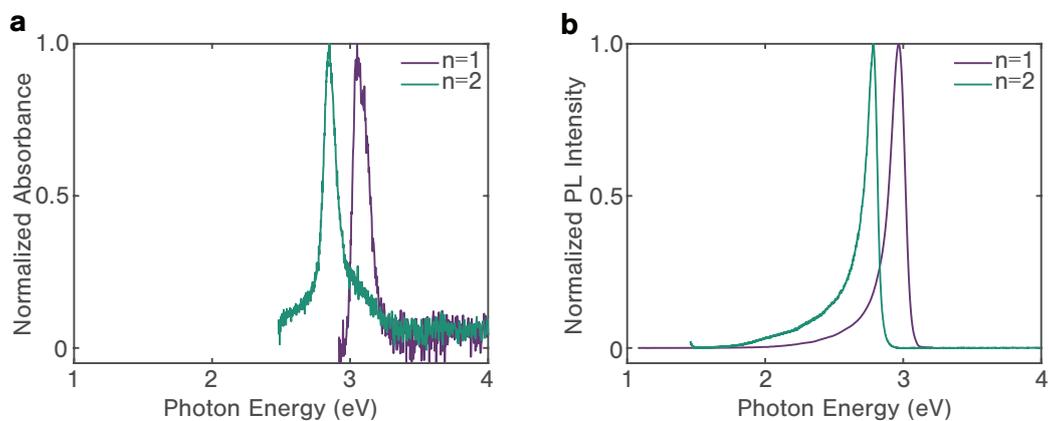

**Fig. S1: Absorption (a) and photoluminescence (b) spectra of n=1 and n=2 2DHPs adapted from Ref [1].**

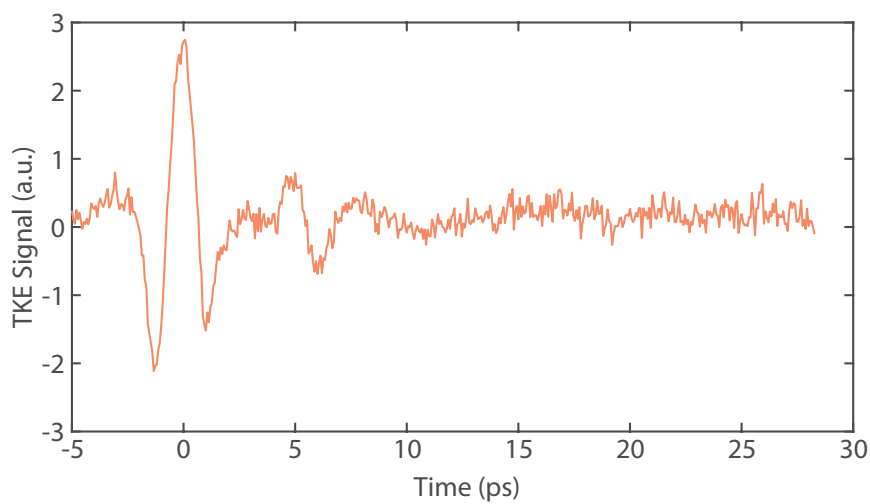

**Fig. S2: TKE signals of n-butylammonium bromide crystal at room temperature.** The initial TKE signal shows a bipolar response similar to that observed in the n=1 sample. The second peak at ∼ 5 ps is due to the reflection of the THz pulse within crystal.

9

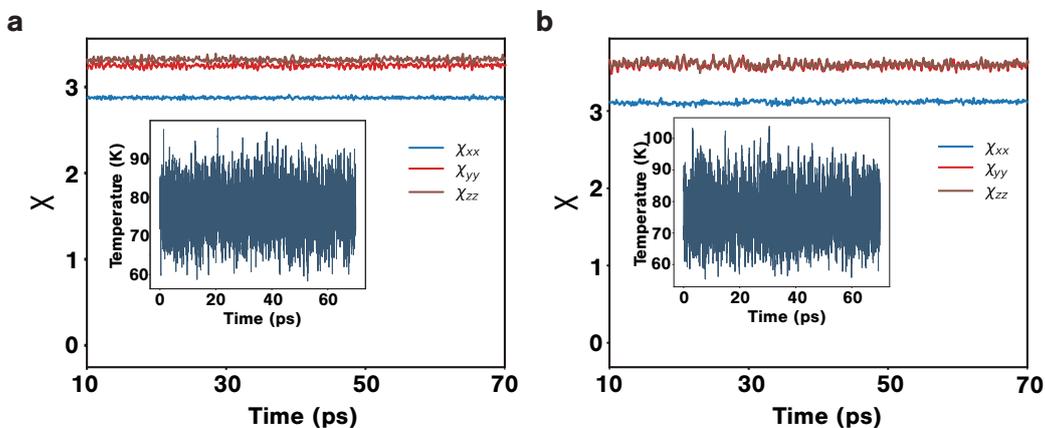

**Fig. S3: Time evolution of susceptibilities of 2DHPs at 77 K.** Non-zero tensor components of n=1 (**a**) and n=2 (**b**) of time-dependent susceptibilities sampled from finite-temperature trajectories at every 100 fs. Inset plots show the corresponding thermal fluctuations during the same time interval.

Table S1: Reagent quantities used for bromide 2D LHP syntheses

| Items | $(BA)_2PbBr_4$ | $(BA)_2MAPb_2Br_7$ |
|---|---|---|
| PbO mass (g) | 0.558 | 0.279 |
| Number of moles of PbO (m*mol*) | 2.5 | 1.25 |
| HBr volume to make $PbBr_2$ solution (mL) | 3 | 1.5 |
| MABr mass (g) | - | 0.0700 |
| Number of moles of MABr (m*mol*) | - | 0.625 |
| HBr volume to make MABr solution (mL) | - | 0.4 |
| BA volume ($\mu$L) | 247 | 74 |
| Number of moles of BA (m*mol*) | 2.5 | 0.75 |
| Additional volume of HBr for dilution (mL) | 5 | 0.5 |
| Total volume of HBr used (mL) | 8 | 2.4 |


\end{document}